\documentclass[aps,prl,twocolumn,showpacs,amsmath,amssymb,bm,superscriptaddress]{revtex4-1}  
\usepackage{graphicx}  
\usepackage{dcolumn}   
\usepackage{hyperref}
\usepackage{bm}        

\newcommand{\ajpk}{\ensuremath{A^{J/\psi K}}}
\newcommand{\ajppi}{\ensuremath{A^{J/\psi \pi}}}
\newcommand{\BplusDecay}{\ensuremath{B^+ \rightarrow J/\psi K^{+}}}
\newcommand{\BminusDecay}{\ensuremath{B^- \rightarrow J/\psi K^{-}}}
\newcommand{\BplusPiDecay}{\ensuremath{B^+ \rightarrow J/\psi \pi^{+}}}
\newcommand{\BminusPiDecay}{\ensuremath{B^- \rightarrow J/\psi \pi^{-}}}
\newcommand {\ks} {\ensuremath{K^0_S}}

\begin{document}

\hspace{5.2in} \mbox{Fermilab-Pub-13-088-E}

\title{Measurement of  direct $\bm{CP}$ violation parameters in 
$\bm{B^\pm \rightarrow J/\psi K^{\pm}}$ and $\bm{B^\pm \rightarrow J/\psi \pi^{\pm}}$  decays with  10.4~fb$^{-1}$ of Tevatron data}

\affiliation{LAFEX, Centro Brasileiro de Pesquisas F\'{i}sicas, Rio de Janeiro, Brazil}
\affiliation{Universidade do Estado do Rio de Janeiro, Rio de Janeiro, Brazil}
\affiliation{Universidade Federal do ABC, Santo Andr\'e, Brazil}
\affiliation{University of Science and Technology of China, Hefei, People's Republic of China}
\affiliation{Universidad de los Andes, Bogot\'a, Colombia}
\affiliation{Charles University, Faculty of Mathematics and Physics, Center for Particle Physics, Prague, Czech Republic}
\affiliation{Czech Technical University in Prague, Prague, Czech Republic}
\affiliation{Institute of Physics, Academy of Sciences of the Czech Republic, Prague, Czech Republic}
\affiliation{Universidad San Francisco de Quito, Quito, Ecuador}
\affiliation{LPC, Universit\'e Blaise Pascal, CNRS/IN2P3, Clermont, France}
\affiliation{LPSC, Universit\'e Joseph Fourier Grenoble 1, CNRS/IN2P3, Institut National Polytechnique de Grenoble, Grenoble, France}
\affiliation{CPPM, Aix-Marseille Universit\'e, CNRS/IN2P3, Marseille, France}
\affiliation{LAL, Universit\'e Paris-Sud, CNRS/IN2P3, Orsay, France}
\affiliation{LPNHE, Universit\'es Paris VI and VII, CNRS/IN2P3, Paris, France}
\affiliation{CEA, Irfu, SPP, Saclay, France}
\affiliation{IPHC, Universit\'e de Strasbourg, CNRS/IN2P3, Strasbourg, France}
\affiliation{IPNL, Universit\'e Lyon 1, CNRS/IN2P3, Villeurbanne, France and Universit\'e de Lyon, Lyon, France}
\affiliation{III. Physikalisches Institut A, RWTH Aachen University, Aachen, Germany}
\affiliation{Physikalisches Institut, Universit\"at Freiburg, Freiburg, Germany}
\affiliation{II. Physikalisches Institut, Georg-August-Universit\"at G\"ottingen, G\"ottingen, Germany}
\affiliation{Institut f\"ur Physik, Universit\"at Mainz, Mainz, Germany}
\affiliation{Ludwig-Maximilians-Universit\"at M\"unchen, M\"unchen, Germany}
\affiliation{Panjab University, Chandigarh, India}
\affiliation{Delhi University, Delhi, India}
\affiliation{Tata Institute of Fundamental Research, Mumbai, India}
\affiliation{University College Dublin, Dublin, Ireland}
\affiliation{Korea Detector Laboratory, Korea University, Seoul, Korea}
\affiliation{CINVESTAV, Mexico City, Mexico}
\affiliation{Nikhef, Science Park, Amsterdam, the Netherlands}
\affiliation{Radboud University Nijmegen, Nijmegen, the Netherlands}
\affiliation{Joint Institute for Nuclear Research, Dubna, Russia}
\affiliation{Institute for Theoretical and Experimental Physics, Moscow, Russia}
\affiliation{Moscow State University, Moscow, Russia}
\affiliation{Institute for High Energy Physics, Protvino, Russia}
\affiliation{Petersburg Nuclear Physics Institute, St. Petersburg, Russia}
\affiliation{Instituci\'{o} Catalana de Recerca i Estudis Avan\c{c}ats (ICREA) and Institut de F\'{i}sica d'Altes Energies (IFAE), Barcelona, Spain}
\affiliation{Uppsala University, Uppsala, Sweden}
\affiliation{Lancaster University, Lancaster LA1 4YB, United Kingdom}
\affiliation{Imperial College London, London SW7 2AZ, United Kingdom}
\affiliation{The University of Manchester, Manchester M13 9PL, United Kingdom}
\affiliation{University of Arizona, Tucson, Arizona 85721, USA}
\affiliation{University of California Riverside, Riverside, California 92521, USA}
\affiliation{Florida State University, Tallahassee, Florida 32306, USA}
\affiliation{Fermi National Accelerator Laboratory, Batavia, Illinois 60510, USA}
\affiliation{University of Illinois at Chicago, Chicago, Illinois 60607, USA}
\affiliation{Northern Illinois University, DeKalb, Illinois 60115, USA}
\affiliation{Northwestern University, Evanston, Illinois 60208, USA}
\affiliation{Indiana University, Bloomington, Indiana 47405, USA}
\affiliation{Purdue University Calumet, Hammond, Indiana 46323, USA}
\affiliation{University of Notre Dame, Notre Dame, Indiana 46556, USA}
\affiliation{Iowa State University, Ames, Iowa 50011, USA}
\affiliation{University of Kansas, Lawrence, Kansas 66045, USA}
\affiliation{Louisiana Tech University, Ruston, Louisiana 71272, USA}
\affiliation{Northeastern University, Boston, Massachusetts 02115, USA}
\affiliation{University of Michigan, Ann Arbor, Michigan 48109, USA}
\affiliation{Michigan State University, East Lansing, Michigan 48824, USA}
\affiliation{University of Mississippi, University, Mississippi 38677, USA}
\affiliation{University of Nebraska, Lincoln, Nebraska 68588, USA}
\affiliation{Rutgers University, Piscataway, New Jersey 08855, USA}
\affiliation{Princeton University, Princeton, New Jersey 08544, USA}
\affiliation{State University of New York, Buffalo, New York 14260, USA}
\affiliation{University of Rochester, Rochester, New York 14627, USA}
\affiliation{State University of New York, Stony Brook, New York 11794, USA}
\affiliation{Brookhaven National Laboratory, Upton, New York 11973, USA}
\affiliation{Langston University, Langston, Oklahoma 73050, USA}
\affiliation{University of Oklahoma, Norman, Oklahoma 73019, USA}
\affiliation{Oklahoma State University, Stillwater, Oklahoma 74078, USA}
\affiliation{Brown University, Providence, Rhode Island 02912, USA}
\affiliation{University of Texas, Arlington, Texas 76019, USA}
\affiliation{Southern Methodist University, Dallas, Texas 75275, USA}
\affiliation{Rice University, Houston, Texas 77005, USA}
\affiliation{University of Virginia, Charlottesville, Virginia 22904, USA}
\affiliation{University of Washington, Seattle, Washington 98195, USA}
\author{V.M.~Abazov} \affiliation{Joint Institute for Nuclear Research, Dubna, Russia}
\author{B.~Abbott} \affiliation{University of Oklahoma, Norman, Oklahoma 73019, USA}
\author{B.S.~Acharya} \affiliation{Tata Institute of Fundamental Research, Mumbai, India}
\author{M.~Adams} \affiliation{University of Illinois at Chicago, Chicago, Illinois 60607, USA}
\author{T.~Adams} \affiliation{Florida State University, Tallahassee, Florida 32306, USA}
\author{J.P.~Agnew} \affiliation{The University of Manchester, Manchester M13 9PL, United Kingdom}
\author{G.D.~Alexeev} \affiliation{Joint Institute for Nuclear Research, Dubna, Russia}
\author{G.~Alkhazov} \affiliation{Petersburg Nuclear Physics Institute, St. Petersburg, Russia}
\author{A.~Alton$^{a}$} \affiliation{University of Michigan, Ann Arbor, Michigan 48109, USA}
\author{A.~Askew} \affiliation{Florida State University, Tallahassee, Florida 32306, USA}
% ...
%
%
\collaboration{The D0 Collaboration\footnote{with visitors from
$^{a}$Augustana College, Sioux Falls, SD, USA,
$^{b}$The University of Liverpool, Liverpool, UK,
$^{c}$DESY, Hamburg, Germany,
$^{d}$Universidad Michoacana de San Nicolas de Hidalgo, Morelia, Mexico
$^{e}$SLAC, Menlo Park, CA, USA,
$^{f}$University College London, London, UK,
$^{g}$Centro de Investigacion en Computacion - IPN, Mexico City, Mexico,
$^{h}$Universidade Estadual Paulista, S\~ao Paulo, Brazil,
$^{i}$Karlsruher Institut f\"ur Technologie (KIT) - Steinbuch Centre for Computing (SCC)
and
$^{j}$Office of Science, U.S. Department of Energy, Washington, D.C. 20585, USA.
}} \noaffiliation
\vskip 0.25cm

\date{5 April 2013, revised manuscript submitted 20 May 2013}

\begin{abstract}
We present a  measurement of the direct $CP$-violating charge asymmetry in $B^\pm$  mesons  
decaying  to $J/\psi K^{\pm}$ and $J/\psi \pi^{\pm}$ where $J/\psi$ decays to $\mu^+ \mu^-$, 
using the full Run II data set of 10.4 fb$^{-1}$ of proton-antiproton collisions collected using the D0 detector 
at the Fermilab Tevatron Collider. A difference in the yield of $B^-$ and  $B^+$ mesons in these decays 
is found by fitting to the difference between their reconstructed invariant mass distributions 
resulting in asymmetries of $\ajpk =\left[ \rm{0.59} \pm 0.37 \right]\%$, 
which is  the most precise measurement to date, and $\ajppi= \left[ \rm{-4.2} \pm  4.5 \right]\%$.
Both measurements are consistent with  standard model predictions.
\end{abstract}

\maketitle

Currently all measurements of $CP$ violation, either in decay, mixing, or in the interference 
between the two, have been consistent with the presence of a single phase in the CKM matrix. 
The standard model predicts that for $b \rightarrow sc\bar{c}$ decays, the tree and penguin 
contributions have the same weak phase, and thus no direct $CP$ violation is expected in the 
decays of $B^\pm$ mesons to $J/\psi K^{\pm}$. 
Estimates of the effect of penguin loops~\cite{hou} show that there could be a small amount of 
direct $CP$ violation of up to ${\cal O}(0.3\%)$.
A measurement of a relatively large charge asymmetry would indicate the existence 
of physics beyond the standard model~\cite{hou, barger, wu}.
In the transition $b \rightarrow dc\bar{c}$, the tree and penguin contributions have 
different phases, and  there may be measurable levels of $CP$ violation in the 
decay $B^\pm \rightarrow J/\psi \pi^{\pm}$~\cite{bscc,hou2}.

The $CP$-violating charge asymmetry in the decays $B^\pm \rightarrow J/\psi K^{\pm}$ 
and $B^\pm \rightarrow J/\psi \pi^{\pm}$ are defined as 
\begin{align}
\ajpk  =  &\frac{\Gamma\left(\BminusDecay \right) - \Gamma\left(\BplusDecay \right)}{\Gamma\left(\BminusDecay \right) + \Gamma\left(\BplusDecay \right)},\\
\ajppi =  &\frac{\Gamma\left(\BminusPiDecay \right) - \Gamma\left(\BplusPiDecay \right)}{\Gamma\left(\BminusPiDecay \right) + \Gamma\left(\BplusPiDecay \right)}. 
\end{align}
Previous measurements of \ajpk \cite{belle2010,d02008,belle2008,babar2005,cleo2000} have been 
averaged by the Particle Data Group with the result 
$\ajpk = \left[  0.1 \pm 0.7 \right]\%$~\cite{pdg2012}. The most precise measurement of \ajpk\ 
was made by the Belle collaboration~\cite{belle2010}, with a total uncertainty of $0.54\%$. 
The most precise measurement of \ajppi\ was made by the LHCb collaboration~\cite{lhcb2012}, 
with a total uncertainty of $2.9\%$. The LHCb measurement is actually a measurement of the 
difference, $\ajppi - \ajpk$, and assumes that \ajpk\ is zero. The previous measurement made 
by the D0 Collaboration~\cite{d02008} has a total uncertainty of $0.68\%$ for \ajpk\ and 
$8.5\%$ for \ajppi\ using a data sample of 2.8  fb$^{-1}$ of proton-antiproton collisions.

This Letter presents substantially improved measurements of \ajpk\ and \ajppi\ using 
the full Tevatron Run II data sample with an integrated luminosity of 10.4~fb$^{-1}$. 
We assume there is no production asymmetry between $B^+$ and $B^-$ mesons 
in proton-antiproton collisions. An advantage of these decay modes into $J/\psi X^\pm$ 
is that no assumptions on the $CP$ symmetry of subsequent charm decays need to 
be made.

These updated measurements of \ajpk\ and \ajppi\ make use of the methods for extracting 
asymmetries used in the analyses of the time-integrated flavor-specific semileptonic charge 
asymmetry in the decays of neutral $B$ mesons~\cite{d0assl,d0adsl}. We measure the 
raw asymmetries 
\begin{align}
A_{\rm raw}^{J/\psi K}  = &\frac{N_{J/\psi K^-} - N_{J/\psi K^+}  }{N_{J/\psi K^-} + N_{J/\psi K^+} },\\
A_{\rm raw}^{J/\psi \pi}  = &\frac{N_{J/\psi \pi^-} - N_{J/\psi \pi^+}  }{N_{J/\psi \pi^-} + N_{J/\psi \pi^+} },
\end{align}
where $N_{J/\psi K^-}$ ($N_{J/\psi K^+}$) is the number of reconstructed \BminusDecay\ 
(\BplusDecay ) decays, and $N_{J/\psi \pi^-}$ ($N_{J/\psi \pi^+}$) is the number of 
reconstructed \BminusPiDecay\ (\BplusPiDecay ) decays. The charge asymmetry in $B^\pm$ 
decays is then given by (neglecting any terms second-order or higher in the asymmetry) 
\begin{align}
\label{eq:asymm}
A^{J/\psi K}  = & A_{\rm raw}^{J/\psi K}  + A_{K},\\
A^{J/\psi \pi}  = & A_{\rm raw}^{J/\psi \pi} + A_\pi,
\end{align}
where $A_K$ is the dominant correction and is the reconstruction asymmetry between 
positively, $\epsilon({K^+})$, and negatively, $\epsilon({K^-})$, charged kaons in 
the detector~\cite{d0det}:
\begin{align}
\label{eq:aK}
A_K = & \frac{\epsilon({K^+}) - \epsilon(K^-)}{\epsilon({K^+}) + \epsilon(K^-)}.
\end{align}
The correction $A_K$ is calculated using the measured kaon reconstruction asymmetry 
as described below~\cite{d0adsl}. As discussed later, data collected using regular 
reversals of magnet polarities results in no significant residual track reconstruction
asymmetries, and hence, no correction for tracking asymmetries or pion reconstruction 
asymmetries need to be applied, hence $A_\pi =0$.

The D0 detector has a central tracking system, consisting of a silicon microstrip tracker  
and the central fiber tracker, both located within a 2~T superconducting solenoidal 
magnet~\cite{d0det, layer0}. A muon system, covering $|\eta|<2$~\cite{eta}, 
consists of a layer of tracking detectors and scintillation trigger 
counters in front of 1.8~T toroidal magnets, followed by two similar layers 
after the toroids~\cite{run2muon}. 

The polarities of the toroidal and solenoidal magnetic fields are reversed
on average every two weeks so that the four solenoid-toroid polarity
combinations are exposed to approximately the same
integrated luminosity. This allows for a cancelation of first-order
effects related to instrumental asymmetries.
To ensure optimal cancelation of the uncertainties, the events are weighted 
according to the number of $J/\psi h^\pm$ decays for each data sample corresponding 
to a different configuration of the magnets' polarities (polarity-weighting). 
The weighting is based on the number of events that pass the selection criteria 
and the likelihood selection (described below) and that are in the $J/\psi h^\pm$ 
invariant mass range used to fit the data.

The data are collected with a suite of single and dimuon triggers. 
The selection and  reconstruction of  $J/\psi h^\pm$ events where 
$h^\pm$ is any stable charged hadron and  $J/\psi \rightarrow \mu^+ \mu^-$ 
requires three tracks with at least two hits in both the silicon 
microstrip tracker  and the central fiber tracker. 
The muon selection requires a transverse momentum $p_T > 1.5$~GeV/$c$ with 
respect to the beam axis. One of the reconstructed muons is required 
to have hits in at least two layers of the muon system
with segments reconstructed both  inside and outside the toroid.  
The second muon is required to have hits in at least the first layer 
of the muon system. The muon track segment has to be matched to a particle 
found in the central tracking system. The dimuon system must have a 
reconstructed invariant mass between 2.80 and 3.35~GeV/$c^2$ consistent 
with the $J/\psi$ mass, $3.097$~GeV/$c^2$ \cite{pdg2012}.

An additional charged particle with $p_T > 0.7$~GeV$/c$ is selected. 
Since the D0 detector is unable to distinguish between $K^\pm$ and $\pi^\pm$, 
and since the $J\psi K^\pm$ process is dominant, this particle is assigned 
the charged  kaon mass and is required to be consistent with coming from the 
same three-dimensional vertex as the two muons, with the $\chi^2$ of the
vertex fit being less than 16 for 3 degrees of freedom. The
displacement of this vertex from the primary proton-antiproton interaction
point is required to exceed 3 standard deviations for the resolution of 
the vertex position in the plane perpendicular to the beam direction. 

The $B^\pm$ selection is further improved using a likelihood ratio method
taken directly from Refs.~\cite{williams, holubyev,d02007,like_ratio} 
that combines a number of variables to discriminate between signal and 
background: the smaller of the transverse momenta of the two muons;
the $\chi^2$ of the $B$ decay vertex;
the $B^\pm$ decay length divided by its uncertainty; 
the significance, $S_B$, of the reconstructed $B^\pm$ meson impact parameter;  
the transverse momentum of the $h^\pm$;
and the significance, $S_K$, of the $h^\pm$  impact parameter.

For any particle $i$, the significance $S_i$ is defined as 
$S_i = \sqrt{[\epsilon_T/\sigma(\epsilon_T)]^2 + 
[\epsilon_L/\sigma(\epsilon_L)]^2}$, where $\epsilon_T$ ($\epsilon_L$)
is the projection of the impact parameter on the plane
perpendicular to (along) the beam direction,
and $\sigma(\epsilon_T)$ [$\sigma(\epsilon_L)$] is its uncertainty.
The trajectory of each $B^\pm$ is formed assuming that 
it passes through the reconstructed $B^\pm$ vertex and is directed 
along the reconstructed $B^\pm$ momentum.

The final requirement on the likelihood ratio variable is chosen to minimize the 
statistical uncertainty on  $A_{\rm raw}^{J/\psi K}$.
The measurement of  $A_{\rm raw}^{J/\psi \pi}$ makes use of a different selection 
on the likelihood ratio  that minimizes the statistical uncertainty of 
$A_{\rm raw}^{J/\psi \pi}$. The asymmetry results extracted with both of these
likelihood selections are consistent.  No event has more than one possible track 
and $J/\psi$ mass combination that passes all of the selection criteria. 

From each set of three particles fulfilling these requirements, a $J/\psi h^\pm$ 
candidate is constructed. The momenta of the muons are corrected by constraining 
the $J/\psi$ mass to the world average~\cite{pdg2012}.

\begin{figure}[htbp]
\includegraphics[width=\columnwidth]{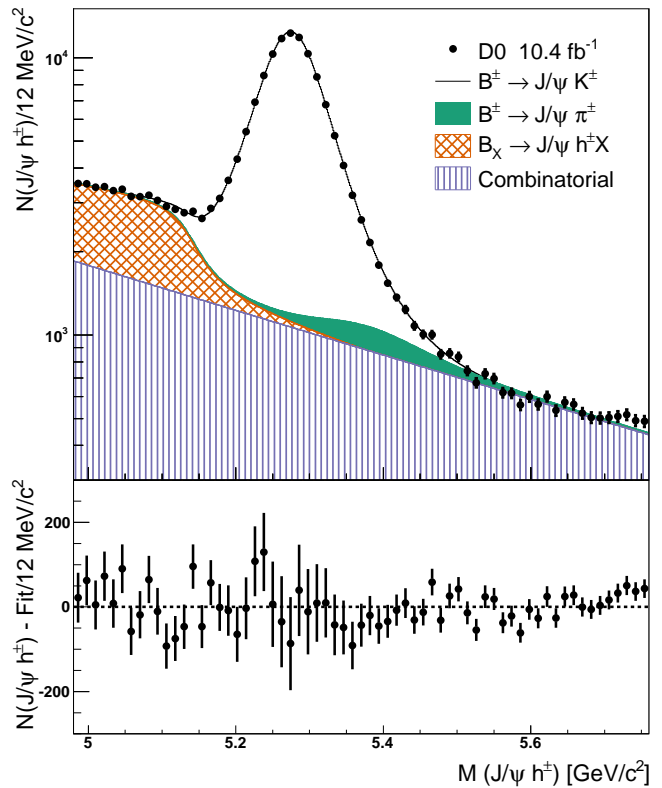}
\caption{\label{Fig:SumFit} 
The polarity-weighted $J/\psi h^\pm$ invariant mass distribution, where the $h^\pm$ is  
assigned the charged kaon mass, after selecting on 
the likelihood-ratio function optimized for $A_{\rm raw}^{J/\psi K}$. The bottom panel shows the  fit residuals (the error bars represent the statistical uncertainty). 
Fit described in the text.
}
\end{figure}

The number of signal candidates are extracted from the $J/\psi h^{\pm}$ mass 
distribution using an unbinned  maximum likelihood fit over a mass range 
of $4.98 <  M(J/\psi h^{\pm}) < 5.76$~GeV$/c^2$ as shown in Fig.~\ref{Fig:SumFit}.
The dominant peak consists of the overlap of the $B^\pm \rightarrow J/\psi K^\pm$ 
and the $B^\pm \rightarrow J/\psi \pi^\pm$ (where the $\pi^\pm$ is mis-identified 
as a $K^\pm$) components. The mis-identified $B^\pm \rightarrow J/\psi \pi^\pm$ 
decay mode appears as a small peak shifted to a slightly higher mass than the $B^\pm$.
The $B^\pm \rightarrow J/\psi K^\pm$ signal peak is modeled by two Gaussian functions 
constrained to have the same mean but, with different widths and normalizations to 
model the detector's mass resolution, $G_K(m)$. Taking account the D0 momentum scale, 
the mean is found to be consistent with the PDG average of the $B^\pm$ meson mass.
To obtain a good fit to the data, the widths  
have a linear dependence on the kaon energy. 
We assume that the mass distribution of the $B^\pm \rightarrow J/\psi \pi^\pm$ 
is identical to that of $B^\pm \rightarrow J/\psi K^\pm$, if the correct hadron 
mass is assigned. To  model the $J/\psi \pi^\pm$  mass distribution, $G_\pi(m)$, 
the $J/\psi \pi^\pm$ signal peak is transformed by assigning the pion track the 
charged kaon mass.
Partially reconstructed decays such as $B_x \rightarrow J/\psi h^\pm X$ where 
$h^\pm$ is any stable charged  hadron and $X$ is additional charged or neutral 
particles (e.g.,~the decay $B^\pm \rightarrow J/\psi  K^{\ast\pm}$) can be 
empirically modeled with a threshold function that extends to  the $B^\pm$ mass and 
is based on Monte Carlo simulations~\cite{holubyev}: 
$T(m) = \arctan\left[p_1  (m c^2 - p_2 ) \right] + p_3$, where $p_i$ are fit parameters.
In the default fit only the normalization of $T(m)$ is allowed to vary and 
the other parameters are fixed to the values obtained from simulation.
The combinatorial background is described by an exponential function, $E(m)$, 
with a slope that depends on the kaon energy.
The fractions of the $J/\psi K$, $J/\psi \pi$, and partially reconstructed decays 
depend on the $h^\pm$ momentum. Empirical studies of the data show that this 
dependence can be modeled by the same polynomial function with different scaling 
factors for the $J/\psi K$, $J/\psi \pi$, and partially reconstructed fractions. 
The coefficients of the polynomial and the scaling factors are determined from the fit. 

The likelihood function is defined to simultaneously fit the raw asymmetries, 
$A_{\rm raw}^{J/\psi K(\pi)} $, the asymmetry of the partially reconstructed 
decays, $A_T$,  and  the asymmetry in the combinatorial background, $A_E$:
\begin{align}
 {\cal L} = &  \left(1 - q_h A_{\rm raw}^{J/\psi K} \right) G_{K}(m)  
 + \left(1 - q_h A_{\rm raw}^{J/\psi \pi} \right) G_{\pi}(m)  \nonumber \\
 &  + \left(1 - q_h A_T \right) T(m) 
 + \left(1 - q_h A_E \right) E(m),
\end{align}
where $q_h$ is the charge of the hadron.

The raw asymmetries are extracted by fitting the resulting data sample using the 
unbinned maximum likelihood fit described above. The resulting $J/\psi h^\pm$ polarity-weighted 
invariant mass distribution is shown in Fig.~\ref{Fig:SumFit}.  
The $B^\pm \rightarrow J/\psi K^\pm$ signal contains $105562 \pm 370 \thinspace (\mbox{stat})$ events,  
and the $B^\pm \rightarrow J/\psi \pi^\pm$ signal contains $3110 \pm 174 \thinspace (\mbox{stat})$ events.

The quality of the fit is estimated by projecting the resulting unbinned likelihood fit onto 
the $J/\psi K^\pm$ invariant mass distribution (65 bins in total). A $\chi^2$ is then 
calculated with a value of 76.2 for  47 degrees of freedom (the number of bins less the 
number of fit parameters excluding the asymmetry parameters).

\begin{figure}[hbtp]
\includegraphics[width=\columnwidth]{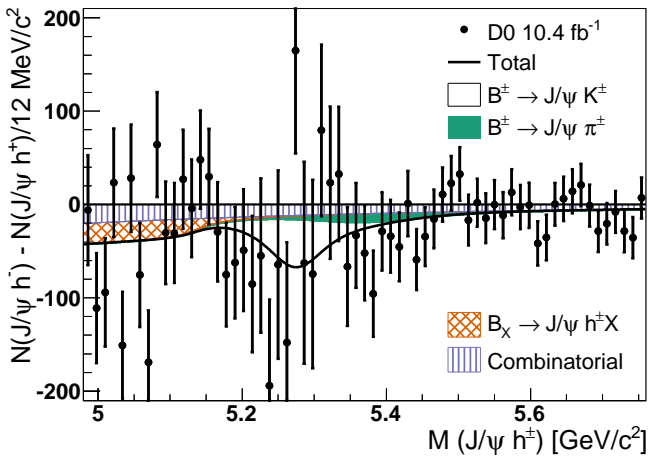}
\caption{\label{Fig:Diff} 
The fit to the difference distribution for the data optimized for 
$A_{\rm raw}^{J/\psi K}$ (the fit is described in the text). 
}
\end{figure}

The invariant mass distribution of the differences, $N(J/\psi h^-) - N(J/\psi h^+)$, 
is  shown in Fig.~\ref{Fig:Diff} with a resulting $\chi^2$ of 58.5 for 61 degrees of 
freedom. The resulting raw asymmetries are extracted from the data are:
\begin{align}
   A_{\rm raw}^{J/\psi K} = {}&  \left[ -0.46 \pm 0.36 \thinspace (\mbox{stat}) \right]\%, \\
   A_{\rm raw}^{J/\psi \pi} = {}&  \left[ -4.2 \pm 4.4 \thinspace (\mbox{stat}) \right]\%.
\end{align}
The  background asymmetries are also determined:
    $A_{T} =  \left[ -1.3 \pm 1.0 \thinspace (\mbox{stat}) \right]\%$
and $A_{E} =  \left[ -1.1 \pm 0.6 \thinspace (\mbox{stat}) \right]\%$.

The systematic uncertainties in the fitting method are evaluated by varying the fitting 
procedure. For each of the following variations the systematic uncertainty is assigned to 
be half the maximum variation in the central value. The mass range of the fit is modified 
so that the lower edge  is varied from 4.95 to 5.01~GeV$/c^2$, and the upper edge 
from 5.73 to 5.79~GeV$/c^2$, in 10~MeV$/c^2$ steps. This results in an uncertainty 
in $A_{\rm raw}^{J/\psi K}$ of $0.022\%$ and in $A_{\rm raw}^{J/\psi \pi}$ of $0.55\%$ 
(labeled ``Mass range'' in Table~\ref{Systematics}).
The following modifications are made to the functions used to model the data. 
The mean of the Gaussian functions is allowed to depend linearly on the 
energy of the kaon. The $p_T(K)$-dependence of the width  of the Gaussian function 
is modeled with a quadratic and a cubic polynomial. The parameters of the threshold 
function are allowed to float. The ratio of branching fractions for the decays 
$B^\pm \rightarrow J/\psi K^\pm$ and $B^\pm \rightarrow J/\psi \pi^\pm$ which 
are not constrained in the default fit are fixed to  the current ratio from 
the Particle Data Group, $0.0482$~\cite{pdg2012}, and the latest measurement by 
the LHCb experiment, $0.0381$~\cite{lhcb2012}. This results in an uncertainty in 
$A_{\rm raw}^{J/\psi K}$ of $0.011\%$ and in $A_{\rm raw}^{J/\psi \pi}$ of 
$0.69\%$ (labelled ``Fit function''). The effect of the event weighting is 
studied by varying the number of events for each magnet configuration by the 
statistical uncertainty ($\sqrt{N}$). This results in  uncertainties of less 
than $0.0005\%$ in \ajpk\ and $0.014\%$ in \ajppi , which are small compared 
to the other uncertainties and is not included in the summary table.

The resulting systematic uncertainties are added in quadrature to obtain:
\begin{align}
   A_{\rm raw}^{J/\psi K} = {}&  \left[ -0.46 \pm 0.36 \thinspace (\mbox{stat}) \pm 0.025 \thinspace (\mbox{syst}) \right]\%, \\
   A_{\rm raw}^{J/\psi \pi} = {}&  \left[ -4.2 \pm 4.4 \thinspace (\mbox{stat}) \pm 0.88 \thinspace (\mbox{syst}) \right]\%.
\end{align}

As a cross-check the following variations of the various asymmetry models are also examined. 
The asymmetries representing the threshold function and the combinatoric 
background are set to the same value, $A_T = A_E$. The asymmetry of the 
combinatoric background is set to zero, $A_E = 0$. The asymmetry of
the threshold function is set to zero, $A_T=0$. The asymmetries representing 
the threshold function and the combinatoric background are both set to zero, 
$A_E = A_T = 0$. When extracting $A_{\rm raw}^{J/\psi K}$, the asymmetry 
$A_{\rm raw}^{J/\psi \pi}$ is set equal to zero. When extracting $A_{\rm raw}^{J/\psi \pi}$, 
the asymmetry $A_{\rm raw}^{J/\psi K}$ is set equal to zero. This results in  
variations in $A_{\rm raw}^{J/\psi K}$ of $0.038\%$ and in $A_{\rm raw}^{J/\psi \pi}$ 
of $1.59\%$. Given the statistical and systematic uncertainties, the observed variations are consistent with no significant biases.

To test the sensitivity of the fitting procedure, the charge of the charged 
hadron in the data is randomized to produce samples with no asymmetry, and  
1000 trials are performed, each with the same statistics as the measurement.
The central value of the asymmetry distribution, $\left(+0.008 \pm 0.011\right)\%$,  
is consistent with zero with a  width of $0.37\%$, consistent with the statistical 
uncertainty found in data. These studies are repeated with introduced asymmetries
of $-1.0$, $-0.5$ and $1.0\%$, each of which returns the expected asymmetries and
statistical uncertainties with no significant bias. 

The residual detector tracking asymmetry has been 
studied in Ref.~\cite{d0assl, d0adsl,dimuon2010} using $\ks \rightarrow \pi^+\pi^-$ 
and $K^{\ast\pm} \rightarrow \ks \pi^\pm$ decays. No significant residual track 
reconstruction asymmetries are found and no correction for tracking asymmetries 
need to be  applied. The tracking asymmetry of charged pions has been studied 
using MC simulations of the detector. The asymmetry is found to be less than $0.05\%$, 
which is assigned as a systematic uncertainty (labeled $\Delta A_{\rm{tracking}}$). 

The correction $A_K$ (Eq.~\ref{eq:aK}), is calculated using the measured kaon 
reconstruction asymmetry presented in Ref.~\cite{d0adsl}.
Negative kaons can interact with matter to produce hyperons, 
while there is no equivalent interaction for positive kaons. 
As a result, the mean path length for positive kaons is larger, 
the reconstruction efficiency is higher, and the kaon asymmetry, $A_K$, is positive. 

The kaon asymmetry is measured using a dedicated sample of 
$K^{\ast 0} (\bar{K}^{\ast 0}) \to K^+\pi^- (K^- \pi^+)$ decays,
based on the technique described in Ref.~\cite{dimuon2010}. 
The $K^+\pi^-$ and $K^-\pi^+$ signal yields are extracted by 
fitting the charge-specific $M(K^{\pm}\pi^{\mp})$ distributions, 
and the asymmetry is determined by dividing the difference by the sum.
The track selection criteria are the same as those required for the 
$J/\psi h^\pm$ signal. 

As expected, an overall positive kaon asymmetry of approximately 1\%  is observed. 
A strong dependence on kaon momentum and the absolute value of the 
pseudorapidity is found, and hence the final kaon asymmetry correction to 
be applied in Eq.~\ref{eq:asymm} is determined by the polarity-weighted average 
of $A_K[p(K),|\eta(K)|]$ over the $p(K)$ and $|\eta(K)|$ distributions in 
the signal events. A relative systematic uncertainty of $5\%$ is 
assigned to each bin to account for possible variations in the yield when different 
models are used to fit the signal and backgrounds in the $K^{*0}$ mass distribution.
Following studies over a range of fit variations, a relative systematic uncertainty of 
3\% on the $J/\psi K^\pm$ yields is applied. The resulting kaon correction is found to 
be (the uncertainty is labeled $\Delta A_K$ in Table~\ref{Systematics}):
\begin{equation}
A_K = \left[ 1.046 \pm 0.043 \thinspace (\mbox{syst})  \right] \%.
\end{equation}
The value of $A_K$ is consistent with that presented in Ref.~7  taking into account  the changes in 
the data selection and the resulting changes in the $p(K)$ and $|\eta(K)|$ distributions.

The final uncertainties are summarized in Table~\ref{Systematics} where 
their combination assumes that they are uncorrelated. We obtain final asymmetries of
\begin{align}
\ajpk =& \left[ \rm{0.59} \pm 0.36 \thinspace(\text{stat}) \pm 0.07 \thinspace(\text{syst}) \right]\%, \label{ajkresult}\\
\ajppi=& \left[ \rm{-4.2} \pm 4.4 \thinspace(\text{stat}) \pm 0.9 \thinspace(\text{syst}) \right]\%. \label{ajpiresult}
\end{align}
This is the most precise measurement of \ajpk\ to date and is a reduction in 
uncertainty by approximately a factor of two from the previous D0 result~\cite{d02008}.

\begin{table}[htbp]
\caption{The statistical and systematic uncertainties  for extracting the asymmetries \ajpk\ and \ajppi .}
\begin{center}
\begin{ruledtabular}
\newcolumntype{A}{D{A}{\pm}{-1}}
\begin{tabular}{cdd}
Type of uncertainty & \multicolumn{1}{c}{\ajpk\ (\%)} &  \multicolumn{1}{c}{\ajppi\ (\%)} \\
\hline
Statistical		 				&	0.36 	& 4.4 	 	\\
\hline 
Mass range 						&	0.022	& 0.55			 \\
Fit function					&	0.011	& 0.69			\\
$\Delta A_{\rm{tracking}}$				&	0.05	& 0.05			\\
$\Delta A_K$							&	0.043	& 	\multicolumn{1}{c}{n/a}	\		\\
\hline 
Total systematic uncertainty 			& 	0.07	& 0.9		\\
\hline 
Total uncertainty 				&   0.37	& 4.5\\
\end{tabular}	
\end{ruledtabular}
\end{center}
\label{Systematics}
\end{table}%

Several consistency checks are performed by dividing the data into smaller samples 
using additional selections based on data-taking periods, magnet polarities, 
transverse momentum, and rapidity of the charged track representing the kaon. 
The resulting variations of \ajpk\ and \ajppi\ are statistically consistent with 
the results of Eqs.~\ref{ajkresult} and \ref{ajpiresult}.

In summary, we have presented the most precise measurement to date of the charge asymmetry 
$\ajpk =\left[ \rm{0.59} \pm 0.36 \thinspace (\text{stat}) \pm 0.07 \thinspace (\text{syst}) \right]\%$ 
using 10.4 fb$^{-1}$ of data. In addition we have improved our measurement of
$\ajppi= \left[ \rm{-4.2} \pm 4.4 \thinspace (\text{stat}) \pm 0.9 \thinspace (\text{syst}) \right]\%$.
Both measurements are in agreement with  standard model predictions.

%
We thank the staffs at Fermilab and collaborating institutions,
and acknowledge support from the
DOE and NSF (USA);
CEA and CNRS/IN2P3 (France);
MON, NRC KI and RFBR (Russia);
CNPq, FAPERJ, FAPESP and FUNDUNESP (Brazil);
DAE and DST (India);
Colciencias (Colombia);
CONACyT (Mexico);
NRF (Korea);
FOM (The Netherlands);
STFC and the Royal Society (United Kingdom);
MSMT and GACR (Czech Republic);
BMBF and DFG (Germany);
SFI (Ireland);
The Swedish Research Council (Sweden);
and
CAS and CNSF (China).
%


\begin{thebibliography}{99}
\bibitem{eta}  
$\eta=-\ln[\tan(\theta/2)]\ $ is the pseudorapidity and $\theta$ is
the polar angle between the track momentum and the proton beam direction. $\phi$ is the azimuthal angle of the track.
\end{thebibliography}
\end{document}